\documentstyle[12pt]{article}
\newcommand{\drm}{{\rm d}}
\newcommand{\bi}{\bibitem}
\newcommand{\til}{\tilde}

\newcommand{\Lc}{{\cal L}}

\newcommand{\Fc}{{\cal F}}

\newcommand{\be}{\begin{equation}}
\newcommand{\ee}{\end{equation}}
\newcommand{\bea}{\begin{eqnarray}}
\newcommand{\eea}{\end{eqnarray}}
\newcommand{\beas}{\begin{eqnarray*}}
\newcommand{\eeas}{\end{eqnarray*}}

\newcommand{\ga}{\gamma}

\newcommand{\erm}{{\rm e}}
\newcommand{\gabf}{\mbox{\boldmath $\gamma$}}

\newcommand{\dar}{\dagger}

\newcommand{\sig}{\sigma}

\newcommand{\ria}{\rightarrow}
\newcommand{\longra}{\longrightarrow}
\newcommand{\upa}{\uparrow}

\newcommand{\veps}{\varepsilon}

\newcommand{\1}{1\!\!1}
\newcommand{\lan}{\langle}

\newcommand{\ran}{\rangle}
\newcommand{\Om}{\Omega}
\newcommand{\th}{\theta}

\newcommand{\R}{I\!\!R}

\newcommand{\h}{\hspace*{0.5 cm}}
\newcommand{\dis}{\displaystyle}

\newcommand{\psit}{\tilde{\psi}}
\newcommand{\bt}{\beta}

\newcommand{\cent}{\centerline}
\newcommand{\vs}{\vspace*}

\newcommand{\r}{\rho}
\newcommand{\dpar}{\partial}
\newcommand{\pa}{\dpar}
\newcommand{\zb}{\bar{z}}

\newcommand{\doz}{\dot{z}}

\newcommand{\dozb}{\dot{\bar{z}}}

\newcommand{\dox}{\dot{x}}
\newcommand{\dopi}{\dot{\pi}}
\newcommand{\dopsi}{\dot{\psi}}

\begin{document}

\cent{{\Large\bf SPIN AND ELECTRON STRUCTURE$^{\: \dag}$}}

\vs{1.2cm}

\cent{{Matej PAV\v{S}I\v{C}$^{*}$, \ Erasmo RECAMI$^{**}$, \  Waldyr A.
RODRIGUES Jr.$^{***}$}}

\vs{0.4 cm}

\cent{{\em I.N.F.N., Sezione di Catania, 57 Corsitalia, Catania, Italy.}}

\footnotetext{$^{\: \dag}$ Work partially supported by INFN, CNR, MURST;
by FUNREI, CNPq, FAPESP; and by the Slovenian Ministry of Science and
Technology.}
\footnotetext{$^{*}$ On leave from the J.Stefan Institute; University of
Ljubljana; 61111--Ljubljana; Slovenia.}
\footnotetext{$^{**}$ Also: Department of Applied Mathematics;
State University at Campinas; 13083--Campinas, S.P.; Brazil; \ Department
of Physics; State University at Catania; 95129--Catania; Italy; \ and:
Facolt\`a di Ingegneria; Universit\`a degli studi di Bergamo; Bergamo; Italy.}
\footnotetext{$^{***}$ On leave from Departamento de Matem\'{a}tica Aplicada
 --- Imecc; UNICAMP; 13083--Campinas, S.P.; Brazil.}

\vs{0.5 cm}

\cent{and}

\vs{0.5 cm}

\cent{G. Daniele MACCARRONE, \ Fabio RACITI, \ Giovanni SALESI}

\vs{0.4 cm}

\cent{{\em Dipartimento di Fisica, Universit\`a di Catania, Catania, Italy.}}

\vs{2cm}

{\small\bf ABSTRACT:}  The recent literature shows a renewed interest, with
various independent approaches, in the classical models for spin.
Considering the possible interest of those results, at least for the
electron case, we purpose in this
paper to explore their physical and mathematical meaning, by the natural and
powerful language of Clifford algebras (which, incidentally, will allow us to
unify those different approaches). \  In such models,
the ordinary electron is in general
associated to the mean motion of a point--like ``constituent" $\cal Q$, whose
trajectory is a cylindrical helix. \ We find, in particular,
that the object $\cal Q$ obeys a new, non-linear
Dirac--like equation, such that ---when averaging over an internal cycle
(which corresponds to a
linearization)--- it transforms into the
ordinary Dirac equation (valid, of course, for the electron as a whole).

\vfill

\newpage

{\bf 1.} {\em Introduction -- \ } The possibility of constructing formal
classical models of spin was already realized at least 60 years ago, from
different points of view.$^{1}$ \ In particular, Schr\"odinger's
suggestion$^{2}$ that the electron spin
was related to its Zitterbewegung (zbw) motion has
been investigated by several authors$^{3}$.

\h In ref.$^{4}$, for instance,
one meets even the proposal of models with
clockwise and anticlockwise ``inner motions"  as classical analogues
of quantum relativistic spinning particles
and antiparticles, respectively. \ The use of grassmannian variables in a
classical
lagrangian formulation for spinning particles was proposed, on the contrary,
by Berezin and Marinov$^{5}$ and
by Casalbuoni$^{6}$. \ Pauri,$^{7}$ moreover,  showed how a
relativistic  (or non-relativistic) spin can enter
a classical formulation, in the ordinary phase space [without
resource now to grassmannian quantities]à, just as a consequence of the
algebraic structure of the Poincar\'e (or Galilei) group. In such an
interesting approach, it was found$^{7}$ that a relativistic classical
zbw motion can directly follow as a ``spin effect" from the requirements
of existence of a lagrangian and of a covariant position in Dirac's
instant form of relativistic dynamics. The quantum analogue of those
developments had been studied in ref.$^{8}$.

\h The number  of papers
appeared on the subject of classical models for spin, starting from the
fifties, is so large that it would be difficult to try quoting them here.
 \ A recent approach,
based on a generalization of Dirac non-linear electrodynamics,
where the spin of the electron is identified with the momentum of the
Poynting vector in a soliton solution of that theory, can be found
in ref.$^{9}$.

\h In this paper we choose, by making recourse to the natural and powerful
language of Clifford algebras,  to refer ourselves mainly to
the model by Barut and Zanghi$^{10,11}$ (BZ), \  which relates the spin
(at least in
the case of the electron) to a helical motion. \ Namely, we first recast
the BZ model into the Clifford formalism, in the meantime clarifying
its physical and mathematical meanings; \ then, we quantize that model for
the electron case. \ In particular, we derive from the BZ model a
{\em new} non-linear
equation for the ``spinorial variable" of the model, which  when
linearized reduces to the ordinary Dirac
equation. \ Solutions of this non-linear equation will be discussed
in forthcoming papers$^{12}$.

\vspace*{1.4 cm}

{\bf 2.} {\em Spin and helical motion -- \ }  In the BZ model,$^{10}$
a classical electron is
characterized, besides by the usual pair of conjugate variables
($x^{\mu},p_{\mu}$), by a second pair of conjugate classical
{\em spinor} variables ($z,i\bar{z}$), representing internal degrees of
freedom, which are functions of an invariant time parameter $\tau$, that
when convenient will be identified with the proper time
of the center of mass (CM).  Quantity $z$ was a Dirac spinor, while
${\bar z} \equiv z^{\dar} \gabf^{0}$. \ Barut and Zanghi, then, introduced
for a spinning particle the classical lagrangian [$c = 1$]

\

\hfill{
${\cal{L}}  \; =  \; {1 \over 2} \lambda i ( {\dot{\bar{z}}} z - \bar{z}
\dot{z} )  \;
+  \; p_{\mu} ({\dot{x}}^{\mu} - \bar{z}   \gabf^{\mu} z)  \; +  \;
e A_{\mu}(x) \bar{z} \gabf^{\mu} z \; \; \; ,$
\hfill} (1)

\

where $\lambda$ has the dimension of an action, $\gabf^{\mu}$ are the Dirac
matrices, and the particle velocity is

\

\hfill{
$v_{\mu} \equiv  \bar{z} \gabf_{\mu} z \; \; .$
\hfill} (1')

\

We are not  writing  down explicitly the spinorial indices of $\bar{z}$ and
$z$. \ Let us consider the simple case of a {\em free} electron
($A_{\mu} = 0$). Then a possible solution of the equations of motion
corresponding to the lagrangian (1) is:$^{\#1}$
\footnotetext{$^{\#1}$ For other solutions, see refs.$^{12}$.}

\

\hfill{$  z(\tau) \; \;  =  \; \; [\cos  m\tau \; - \; i\gabf^{\mu}
\dis{{p_{\mu} \over m}} \sin m\tau] \; z(0) \; \; ,$ \hfill} (2a)

\

\hfill{$  \bar{z}(\tau) \; \;  =  \; \; \bar{z}(0) \; [\cos m\tau \; + \;
i\gabf^{\mu} \dis{{p_{\mu}
\over m}} \sin m\tau] \; \; ,$ \hfill} (2b)

\

and \ $p_{\mu} = {\rm constant}$; \ $p^{2} = m^{2}$; \ $H = p_{\mu} \bar{z}
\gabf^{\mu} z  \equiv p_{\mu}v^{\mu}$; \ and finally:

\

\hfill{$ {\dot{x}}_{\mu}  \; =  \; v_{\mu}  \; =  \; \dis{{p_{\mu} \over
m^{2}}} H  \; +  \; [{\dot{x}}_{\mu}(0) \; - \; \dis{{p_{\mu} \over
m^{2}}} H] \cos 2m\tau \;  +  \; \dis{{{\ddot{x}}_{\mu}(0) \over 2m}}
\sin 2m\tau   $ \hfill} (2c)

\

(which in ref.$^{10}$ appeared with two misprints). In connection with
this ``free" general solution, let us remark that $H$
is a constant of motion so that we can set $H = m$. \ Solution (2) exhibits the
classical analog of the phenomenon known as ``zitterbewegung" (zbw): \ in
fact, the velocity $v_{\mu}
\equiv \dot{x}_{\mu}$ contains the (expected) term $p_{\mu}/m$ plus a
term describing an oscillatory motion with the characteristic frequency
$\omega = 2m$. \ The velocity of the {\em center of mass} will be given by
$W_{\mu} =
p_{\mu}/m$. \ Notice incidentally that, instead of adopting the variables $z$
and $\bar{z}$, one can work in terms of the spin variables, {\em i.e.} in
terms of the dynamical variables $(x_{\mu}, v_{\mu}, \pi_{\mu}, S_{\mu \nu})$,
where:
\

$$v_{\mu} = \dot{x}_{\mu}; \;\; \pi_{\mu} = p_{\mu} - eA_{\mu}; \; \; \; \;
S_{\mu \nu} = {1 \over 4} i \bar{z} [\gabf_{\mu}, \gabf_{\nu}] z \; ,$$

\

so that ${\dot{S}}_{\mu \nu} = \pi_{\mu} v_{\nu} - \pi_{\nu} v_{\mu}$ \ and \
${\dot{v}}_{\mu} = 4 S_{\mu \nu} \pi^{\nu}$. \ In the case of a free electron,
by varying the action corresponding to ${\cal{L}}$ one finds as generator of
space-time rotations the conserved quantity $J_{\mu \nu} = L_{\mu \nu} +
S_{\mu \nu} \;$, \ where \ $S_{\mu \nu}$ is just the particle spin.$^{\#2}$
\footnotetext{$^{\#2}$ Alternatively, Barut and Zanghi,$^{10}$ in order to study
the internal dynamics of the considered (classical) particle, did split
$x_{\mu}$ and $v_{\mu} \equiv {\dot x}_{\mu}$ as follows: \ $x_{\mu} \equiv
X_{\mu} + Q_{\mu}; \; v_{\mu} \equiv W_{\mu} + U_{\mu}$ (where by
definition $W_{\mu} = {\dot X}_{\mu}$ and $U_{\mu} = {\dot Q}_{\mu}$). \ In
the particular case of a
{\em free} particle, ${\dot W}_{\mu} = 0; \; W_{\mu} = p_{\mu}/m$. \ One can
now interpret $X_{\mu}$ and $p_{\mu}$ as the c.m. coordinates, and $Q_{\mu}$
and $P_{\mu} \equiv mU_{\mu}$ as the {\em relative} position and momentum,
respectively. For a free particle, then, one finds that the internal variables
are coordinates oscillating with the zbw frequency $2m$; and that, again, the
total angular momentum $J_{\mu \nu} = L_{\mu \nu} + S_{\mu \nu}$ is a constant
of motion, quantities $S_{\mu \nu}$ being the spin variables.}

\h Let us explicitly observe that solution (2c) is the equation of a
space-time cylindrical helix, {\em i.e.} it represents in 3--space a helical
motion. Let us also stress that this  motion describes
particle spin at a classical level. In fact, such a classical system has
been shown by Barut and Pav\v{s}i\v{c}$^{13}$ to
describe, after quantization, the Dirac electron.$^{13-15}$

\h An alternative approach leading to a classical description of particles
with spin was put forth by Pav\v{s}i\v{c},$^{16-18}$  by making recourse to the
(extrinsic)
curvature of the particle world--line in Minkowski space. \ Let us finally
mention that the {\em same} classical equations of motion
(and the same Poisson--bracket algebra) have been found also in a third
approach, which consists in adding to the ordinary lagrangian an extra term
containing Grassmann variables.$^{19}$

\vspace*{1.4 cm}

{\bf 3.} {\em About the electron structure -- } \  Considering the interest
of the previous results
(which suggest in particular that the helical motion can have a role in the
origin  of spin),
we purpose to explore their physical meaning more deeply, by the very natural
---and powerful--- language of the Clifford algebras:$^{20,21}$ \  in
particular of the ``space-time algebra (STA)" ${\R}_{1,3}$. \ First of all, let
us preliminarily clarify {\em why} Barut and Zanghi had to introduce the Dirac
spinors $z$ in their lagrangian, by recalling that classically the motion of a
spinning top has to be individuated by \ (i) the world--line $\sig$ of its
center of mass ({\em e.g.}, by the ordinary coordinates $x^{\mu}$ and the
conjugate momenta $p_{\mu}$), and \ (ii) a Frenet tetrad$^{\#3}$
\footnotetext{$^{\#3}$ The use of Frenet tetrads in connection with the Dirac
formalism was first investigated in ref.$^{22}$.}
attached$^{22}$ to the world--line $\sig$. \ This continues to be true when
wishing to describe the motion of a point-like spinning particle. \ For the
Frenet tetrad$^{23}$ we have:

\

\hfill{$e_{\mu} \: = \: R \ga_{\mu} {\til{R}} \: = \: \Lambda_{\mu}^{\nu}
\ga_{\nu} \: ; \; \; \; \Lambda_{\mu}^{\nu} \in L_{+}^{\upa}$ \hfill} (3)

\

where $e_{0}$ is parallel to the particle velocity $v$ (even more, $e_0 = v$
whenever one does use as parameter $\tau$ the CM system proper--time);
the tilde represents the {\em reversion}$^{\#4}$;
\footnotetext{$^{\#4}$ \h The main anti-automorphism in $\R_{1,3}$
(called reversion), denoted by the tilde, is such that
$\widetilde{AB} = \til{B} \til{A}$, and $\til{A} = A$ when $A$ is a
scalar or a vector, while $\til{F}=-F$ when $F$ is a 2-vector.}
and $R = R(\tau)$ is a ``Lorentz rotation" [more precisely, \ $R \: \in \:
{\rm Spin}_{+}(1,3)$, and a Lorentz transform of quantity $a$ is
given by $a' = R a \til{R}$]. \ Moreover \ $R \til{R} \: = \:
\til{R} R \: = \: 1 \:$. \ The Clifford STA fundamental
unit--vectors $\ga_{\mu}$  should not be confused with the Dirac
{\em matrices} $\gabf_{\mu}$.
Let us also recall that, while the orthonormal vectors $\ga_{\mu} \equiv
{\pa / {\pa x^{\mu}}}$ constitute a {\em global} tetrad in Minkowski
space-time
(associated with a given inertial observer), on the contrary the Frenet
tetrad $e_{\mu}$ is defined only along $\sig$, in such a way that $e_0$
is tangent to $\sig$. \ At last, it is: \ $\ga^{\mu} =
\eta^{\mu \nu} \ga_{\nu}$, \ and \ $\gamma_5 \equiv \ga_0 \ga_1 \ga_2
\ga_3$.\\
\h Notice that $R(\tau)$ does contain all the essential information carried
by a Dirac spinor. In fact, out of $R$, a ``Dirac--Hestenes" (DH)
spinor$^{24}$ \  $\psi_{\rm DH}$ \ can be constructed as follows:

\

\hfill{$\psi_{\rm DH} \; = \; \r^{1 \over 2} \erm^{{\bt \gamma_{5}} /2} R$
\hfill} (4)

\

where $\r$ is a normalization factor; and \ $\erm^{{\bt \gamma_{5}}} =
+1$ \ for the electron (and $-1$ for the positron); while, if $\veps$ is a
primitive idempotent of the STA, any Dirac spinor $\psi_{\rm D}$ can be
represented in our STA as:$^{25}$

\

\hfill{$\psi_{\rm D} \; = \; \psi_{\rm DH} \, \veps \: .$ \hfill} (5)

\

For instance, the Dirac spinor $z$ introduced by BZ is obtained from the DH
spinor$^{\#5}$ by the choice \ $\veps \: = \: {1 \over 2} (1 + \ga_{0})$. \
\footnotetext{$^{\#5}$ The DH spinors can be regarded as the {\em parent}
spinors, since all the other spinors of common use among physicists
are got from them by operating as in eq.(5). We might
call them ``the {\em fundamental} spinors".}
Incidentally, the Frenet frame can also write \ $\r e_{\mu} \: =
 \: \psi_{\rm DH} \ga_{\mu} {\psit_{\rm DH}}$.\\
\h Let us stress that, to specify how does the Frenet tetrad rotate as $\tau$
varies, one has to single out a particular $R(\tau)$, and therefore a DH
spinor $\psi_{\rm DH}$ , and eventually a Dirac spinor $\psi_{\rm D}$. \ This
makes intuitively clear why the BZ Dirac--spinor $z$ provides a good
description of the ``spin motion" of a classical particle.\\
\h Let us now repeat what precedes  on a more formal ground. \ In the following,
unless differently stated, we shall indicate the DH spinors $\psi_{\rm DH}$
simply by $\psi$.\\

\h  Let us translate the BZ lagrangian into the Clifford language.\\
\h In eq.(1) quantity $z^{\rm T} = (z_1 \;\; z_2 \;\; z_3 \;\; z_4)$ is a
Dirac spinor $\tau \longra z(\tau)$, and $\bar{z} = z^{\dar} \gabf^{0}$.
To perform our translation, we need a matrix representation of the Clifford
space-time algebra; this can be implemented by {\em representing} the
fundamental
Clifford vectors \ ($\ga_0$, $\ga_1$, $\ga_2$, $\ga_3$) \ by the
{\em ordinary} Dirac matrices $\gabf_{\mu}$.  Choosing:

$$\ga_0 \longra \gabf_0 \: \equiv \: \left( \begin{array}{cc} \1 & 0 \\ 0 & -\1
\end{array} \right)
\; ; \;\;\;\;  \ga_i \longra \gabf_i \: \equiv \: \left( \begin{array}{cc}
0 & -\sig_i \\ \sig_i & 0  \end{array} \right) \; ,$$

the representative in  $\R_{1,3}$ of Barut--Zanghi's quantity $z$ is

\

\hfill{$z \longra \Psi \equiv \psi \, \veps \; ; \;\;\; \veps \equiv {1 \over 2}
(1 + \ga_0)$
\hfill} (6)

\

where $\psi$, and $\psit$, are represented \ (with $\psi \psit = \1$) \ by

\

\hfill{$\psi \; = \; \left( \begin{array}{cccc} z_1 & -\zb_2 & z_3 & \zb_4 \\
z_2 & \zb_1 & z_4 & -\zb_3 \\ z_3 & \zb_4 & z_1 & -\zb_2 \\ z_4 & -\zb_3 &
z_2 & \zb_1 \end{array} \right) \; ; \;\;\; \psit \; = \; \left( \begin{array}
{cccc} \zb_1 & \zb_2 & -\zb_3 & -\zb_4 \\
-z_2 & z_1 & -z_4 & z_3 \\ -\zb_3 & -\zb_4 & \zb_1 & -\zb_2 \\ -z_4 & z_3 &
-z_2 & z_1 \end{array} \right) \; .$
\hfill} (7)

\

The translation of the various terms in eq.(1) is then:

\

\hfill{$\begin{array}{lcl} {1 \over 2}i(\dozb z - z\doz) & \longra & \lan
\psit \dopsi \ga_1 \ga_2 \ran_0  \\ p_{\mu}(\dox^{\mu} - \zb \gabf^{\mu} z)
&  \longra & \lan p(\dox - \psi \ga_0 \psit ) \ran_0  \\ e A_{\mu} \zb
\gabf^{\mu} z & \longra & e \: \lan A \psi \ga_0 \psit \ran_0 \; , \end{array}$
\hfill}

\

where $\lan \;\;\; \ran_0$ means ``the scalar part" of the Clifford product.
Thus, the lagrangian $\Lc$ in the Clifford formalism is

\

\hfill{$\Lc \; = \; \lan \psit \dopsi \ga_1 \ga_2 \: + \: p(\dox -
\psi \ga_0 \psit) \: + \: eA \psi \ga_0 \psit \ran_0 \; ,$
\hfill} (8)

\

which is analogous, incidentally, to Kr\"{u}ger's lagrangian$^{26}$ (apart
from a misprint).\\
\h As we are going to see, by ``quantizing" it, also in the present formalism
it is possible (and,
actually, quite easy) to derive from $\Lc$ the Dirac--Hestenes
equation:$^{\#6}$

\

\hfill{$\dpar \, \psi (x) \, \ga_1 \ga_2 \; + \; m \, \psi (x) \,
\ga_0 \; + \; e \, A(x) \, \psi (x) \; = \; 0 \; ,$
\hfill} (9)

\

\footnotetext{$^{\#6}$ Observe that in eq.(8) it is  $\psi = \psi (\tau)$,
while in
eq.(9) we have $\psi = \psi (x)$  with $\psi (x)$ such that its
{\em restriction} \ ${\psi (x)}_{\mid \sig}$ \
to the world--line $\sig$ coincides with $\psi (\tau)$. \ Below, we shall
meet the same
situation, for instance,  when passing from eq.(10a) to eqs.(12)--(12').}
which is nothing but the ordinary Dirac equation written down in the Clifford
formalism.$^{20,24}$ \ Quantity $\dpar \: = \: \ga^{\mu} \dpar_{\mu}$ is the
Dirac operator. \  Let us notice that $p$ in eq.(8) can be regarded as a
Lagrange
multiplier, when the velocity $v = \dox$ is represented by $\psi \ga_0 \psit$.
 \ The BZ model is indeed a hamiltonian system, as proved by using Clifford
algebras in ref.$^{27}$ (cf. also ref.$^{28}$). \
The dynamical variables are then ($\psi , \psit , x, p$), and the
Euler--Lagrange equations yield a system of three independent equations:

\

\hfill{$\dopsi \ga_1 \ga_2 + \pi \psi \ga_0 \; = \; 0 $
\hfill} (10a)

\hfill{$\dox \; = \; \psi \ga_0 \psit $
\hfill} (10b)

\hfill{$\dopi \; = \; e F \cdot \dox $
\hfill} (10c)

\

where \ $F \equiv \pa \wedge A$ \ is the electromagnetic field (a bivector,
in Hestenes' language) \ and \ $\pi \equiv p - eA$ \ is the kinetic
momentum. \ [Notice incidentally, from eq.(10b), that $\dox^2 =
\r (\tau)$].\\
 \h At this point, let us consider a velocity vector {\em field} $V(x)$
together with
its integral lines (or stream--lines).  Be $\sig$ the stream--line along which
a particle moves ({\em i.e.}, the particle world--line). Then, the velocity
distribution $V$ is required to be such that its restriction \
${V(x)}_{\mid \sig}$ \
to the world--line $\sig$ is the ordinary velocity $v =
v(\tau)$ of the considered particle.\\
\h If we moreover recall$^{29,30}$ that any Lorentz ``rotation" $R$
can be written \ $R = \erm^{\Fc}$, \ where $\Fc$ is a {\em bivector},
then along any stream--line $\sig$ we shall have:$^{19}$

\

\hfill{$\dot{R} \equiv \dis{{\drm R \over {\drm \tau}}} \; =
\; {1 \over 2} v^{\mu} \Om_{\mu} R \; = \; {1 \over 2} \Om R \; ,$
\hfill} (11)

\

with \ $\pa_{\mu} R \, = \, \Om_{\mu} R/2$, \ where $\Om_{\mu} \equiv 2
\pa_{\mu} \Fc$, and where $\Om \equiv v^{\mu} \Om_{\mu}$
is the angular--velocity bivector (also known, in differential
geometry, as the ``Darboux bivector"). \ Therefore,
for the tangent vector {\em along any line} $\sig$ we obtain the relevant
relation:

$${\drm \over {\drm \tau}} \; = \; v^{\mu} \pa_{\mu} \; = \; v \cdot \pa \;
$$

The [total derivative] equation (10a) thus becomes:$^{\#6}$

\

\hfill{$v \cdot \pa \psi \ga_1 \ga_2 + \pi \psi \ga_0 \; = \; 0 \; ,$
\hfill} (12)

\

which is a non-linear [partial derivative] equation, as it is easily seen
by using eq.(10b) and rewriting it in the {\em noticeable} form

\

\hfill{$(\psi \ga_0 \psit) \cdot \pa \psi \ga_1 \ga_2 + \pi \psi \ga_0 \; =
\; 0 \; .$
\hfill} (12')

\

Equation (12') constitutes a new {\em non-linear} Dirac--like equation.
 \ The solutions of this new
equation will be explicitly discussed in refs.$^{12}$. \ \
Let us here observe only that the probability current $J \equiv V$ is
conserved: \ $\pa \cdot V = 0$, \ as we shall show elsewhere.

\h Let us pass now to the {\em free} case ($A_{\mu} = 0$), when eq.(10a)
may be written

\

\hfill{$\dopsi \ga_1 \ga_2 + p \psi \ga_0 \; = \; 0 \; ,$
\hfill} (10'a)

\

and admits some simple solutions.$^{12}$ \  Actually, in this case $p$ is
constant [cf. eq.(10c)] and one can choose the {$\ga_{\mu}$} frame so that
 \ $p = m \ga_0$ \ is a constant vector in the direction $\ga_0$. \ Since \
$\dox = \psi \ga_0 \psit$, \ it follows that

\

\hfill{$v \: = \: \dis{{1 \over m}} \psi p \psit \:; \;\;\;\;
\dis{{p \over m}} \: = \: \psi^{-1} v \psit^{-1} \; .$
\hfill} (13)

\

The mean value of $v$ over a zitterbewegung period is then given by the
relation

\

\hfill{$\lan v \ran_{\rm zbw} \: = \: p/m \: = \:
\psi^{-1} v \psit^{-1}$
\hfill} (14)

\

which resembles the ordinary quantum--mechanical mean value for the
wave--function $\psi^{-1}$ (recall that in Clifford algebra for any $\psi$
it exists its inverse). \ Let us recall, by comparison, that the time
average of $v_{\mu}$ given in eq.(2c) over a zbw period is evidently
equal to $p_{\mu}/m$.   \\
\h Let us explicitly stress that, due to the first one [$z \longra \psi \veps$]
of eqs.(6), the results found by BZ for $z$ are valid as well for $\psi$ in
our formalism.  For instance, for BZ [cf. eq.(1')] it was  \ $v^{\mu} =
\zb \gabf^{\mu} z \equiv
v^{\mu}_{\rm BZ}$, \ while in the Clifford formalism [cf. eq.(7)] it is \
$v \: = \: \psi \ga^0 \psi \: = \: \lan \ga^0 \psit \ga^{\mu} \psi \ran_0
\ga_{\mu} \: = \: v^{\mu}_{\rm BZ} \ga_{\mu}$. \ As a consequence, $\sig$
refers in general to a cylindrical helix (for the free case) also in our
formalism.\\
\h Going back to eq.(10'a), by the second one of eqs.(13) we
finally obtain our non-linear (free) Dirac--like equation in the following form:

\

\hfill{$v \cdot \pa \psi \ga_1 \ga_2 \: + \: m \psi^{-1} v \psit^{-1} \psi
\ga_0 \; = \; 0 \; ,$
\hfill} (15)

\

which in the ordinary, tensorial language would write: \ \
$i(\bar{\Psi} \gabf^{\mu} \Psi) \pa_{\mu} \Psi \, = \, \gabf^{\mu} p_{\mu}
\Psi$, \ \
where $\Psi$ and $\gabf^{\mu}$ are now an ordinary Dirac spinor and
the ordinary Dirac matrices, respectively, and \ ${\hat{p}}_{\mu} \equiv
i \pa_{\mu}$.\\

\h In connection with this fundamental equation of motion (15), let us
explicitly notice the following. \ At a classical level, the equation of
motion in the BZ model was eq.(10a), which held
for the world--line $\sig$.  In other words, eq.(10a) was
valid for {\em one} world--line; on the contrary, eq.(15) is
a {\em field} equation, satisfied by quantities $\psi (x)$ such that
\ ${\psi (x)}_{\mid \sig} \, = \, \psi (\tau)$. \ A change in interpretation
is of course necessary when passing from the
classical to the ``quantum" level: and therefore eq.(15) is now to be
regarded as valid
for {\em a congruence} of world lines, that is to say, for a congruence of
stream--lines of the velocity field $V = V(x)$. \ In the quantum case,
the ``particle"
can follow any of those integral lines, with probability amplitude $\rho$.
 \ \ In this context, it must be recalled that a tentative interpretation of
the Dirac equation within the Clifford algebra approach has been suggested
in ref.$^{20}$, and later in ref.$^{31}$. \ However, in the present paper
we shall not put forth,
nor discuss, any interpretation of our formalism.\\
\h As we have seen,  eq.(15) will hold for our helical
motions. Notice moreover that, since [cf. eq.(4)] it is
 \ $\psi = {\r^{1 \over 2}} {\erm^{{\bt \gamma_{5}} /2}} R$, \ in the case
of the helical
path solution the Lorentz ``rotation" $R$ will be the product of a pure space
rotation and a boost.\\

\h It is important to observe that, if we replace $v$ in eq.(15) by its
mean value over a zbw period, eq.(14), then we end up with the
Dirac--Hestenes equation
[{\em i.e.}, the ordinary Dirac equation!], valid now for the center-of-mass
world--line.  In fact, since \ $\lan v \ran_{\rm zbw} \: = \: p/m$, \ eq.(15)
yields

\

\hfill{$p \cdot \pa \psi \ga_1 \ga_2 \: + \: mp \psi \ga_0 \; = \; 0 \; ,$
\hfill} (15')

\

and therefore \ ---if we recall that for the eigenfunctions of $p$ in the DH
approach$^{20}$ it holds \ $\pa \psi \ga_1
\ga_2 = p \psi$, \ so that $(p \cdot \pa) \psi \ga_1 \ga_2 = p \pa \psi \ga_1
\ga_2$--- \ one obtains:

\

\hfill{$p (\pa \psi \ga_1 \ga_2 + m \psi \ga_0) \; = \; 0 \; ,$
\hfill} (15")

\

which is satisfied once it holds the {\em ordinary} Dirac
equation (in its Dirac--Hestenes form):

\

\hfill{$\pa \psi \ga_1 \ga_2 \: + \: m \psi \ga_0 \; = \; 0 \; .$
\hfill} (16)

\

\h Let us observe that all the eigenfunctions of $p$ are solutions both
of eq.(15) and of the Dirac equation.\\

\h In conclusion, our non-linear Dirac--like equation (15) is a
quantum--relativistic equation, that can be regarded as ``sub-microscopic"
in the sense that it refers to the internal motion of a point-like
``constituent" $\cal Q$. In fact, the density current $\psi \ga^0 \psit$,
relative to the solutions $\psi$ of eq.(15), does oscillates in time in
a helical fashion, in
complete analogy to the initial equation (2c); so that $x$ and $\dot{x}$
refer to $\cal Q$. \ \ On the contrary, the ordinary (linear) Dirac
equation can be regarded as the equation describing the global motion of
the geometrical centre
of the system ({\em i.e.}, of the whole ``electron"); actually, it has been
obtained from eq.(15) by linearization, that is to say, replacing the
density current $v \equiv \psi \ga^0 \psit$ by its time
average $p/m$ over the zbw period [cf. eqs.(13)--(14)].\\
\h At last, let us underline that, in the free case,
eq.(10'a) admits also a trivial solution $\sig_0$,
corresponding to rectilinear motion.

\vspace*{1.4 cm}

{\bf 4.} {\em A very simple solution -- } \  In the free case, the equation
(10'a)
admits also a very simple solution (the limit of the ordinary helical
paths when their radius $r$ tends to zero), which ---incidentally--- escaped
BZ's attention; namely:

\

\hfill{$\psi \; = \; \r^{1 \over 2} \exp [- \ga_2 \ga_1 m \tau] \; =
 \; \psi (0) \exp [- \ga_2 \ga_1 m \tau]$
\hfill} (17)

\

as it can be easily verified. \ Notice that expressions (17) are also solutions
of the Dirac equation (16). \ \ Moreover, from eq.(10b) it follows that
 \ \ $v \equiv \dox \; = \; \r \ga_0 $ \ \
and we can set \ $\r = 1$, \ which confirms that the trivial solution (17)
corresponds to rectilinear uniform motion, and that $\tau$ in this case is
just the proper--time along the particle world--line $\sig_0$. \ (In this case,
of course, the zbw disappears).\\

\h Before going on, we want to explicitly put forth the following observation.
Let us first recall
that in our formalism the (Lorenz force) equation of motion for a charged
particle moving with velocity $w$ in an electromagnetic field $F$ is

\

\hfill{${\dot w} \; = \; {\dis{{e \over m}}} F \cdot w \; .$
\hfill} (18)

\

Now, for {\em all} the free--particle solution of the BZ model in the Clifford
language, it holds the ``Darboux relation":

\

\hfill{$ {\dot e}_{\mu} \; = \; \Om \cdot e_{\mu} \; , $
\hfill} (19)

\

so that the ``sub--microscopic" point--like object $\cal Q$, moving along
the helical
path $\sig$, is endowed [cf. eq.(11)] with the angular--velocity bivector

\

\hfill{$\Om \; = \; {1 \over 2} \, {\dot e}_{\mu} \wedge e^{\mu} \; = \;
{1 \over 2} \, {\dot e}_{\mu} e^{\mu} \; ,$
\hfill} (20)

\

as it follows by recalling that $e_{\mu}$ can always be written, like in
eq.(3), as \ $e_{\mu} \: = \: R \ga_{\mu} {\til{R}}$. \ \
Finally, let us observe that eq.(19) yields in particular \ ${\dot e}_{0} \:
= \: \Om \cdot e_{0}$, \ which is formally identical to eq.(18). \ Thus, the
formal, algebraic way we chose [see eq.(18)] for describing that the system as
a whole possesses a non-vanishing magnetic dipole structure suggests
that the bivector field $\Om$ may be regarded as a kind of {\em internal}
electromagnetic--like field, which keeps the ``sub--microscopic" object $\cal Q$
moving along the helix.$^{32-36}$ \  In other words, $\cal Q$ may be
considered as
confining itself along $\sig$ [{\em i.e.}, along a circular orbit, in the
electron C.M.], via the generation of the internal, electromagnetic--like
field

\

\hfill{$F_{\rm int} \; \equiv \; {\dis{{m \over {2 e}}}} \: {\dot e}_{\mu}
\wedge e^{\mu} \; .$
\hfill} \\

\h We shall further discuss this point elsewhere.

\vspace*{1.5 cm}

{\bf 5.} {\em Further remarks -- } \  In connection with our
new equation (12'), or rather with its (free) form (15), we met solutions
corresponding ---in the free case--- to helical motions with constant
radius $r$; as well as a limiting solution, eq.(17), for $r \ria 0$. \
We have seen above
that the latter is a solution also of the ordinary (free) Dirac equation.
Actually, the solution of
the Dirac equation for a free electron in its rest frame can be written$^{20}$
in the present formalism as  \ $\psi (x) \; = \; \psi (0) \exp [- \ga_2 \ga_1
m \tau]$, \
which coincides with our eq.(17) along the world--line $\sig_0$. Let us here
add the observation that,  by recalling that in the Clifford formalism the spin
bivector $S$ is given by \ $ S \; = \; R \ga_2 \ga_1 \til{R} {\dis{\hbar
\over 2}} \; = \; e_2  e_1 {\dis{\hbar \over 2}} \; ,$ \
whilst the angular--velocity bivector $\Om$ is given by eq.(20), one can
finally write

\

\hfill{$p \cdot v \; = \; \Om \cdot S \; = \; m \; ,$
\hfill} (21)

\

which seems to suggest$^{20}$ the electron rest--mass to have an
(internal) kinetic origin!  \ It
is possible, moreover, that such a motion be also at the origin
of electric charge itself; in any case, we already saw that, if the electron is
associated with a clock-wise rotation, then the positron will be associated
with an anti--clock-wise rotation, with respect to the motion direction. \
Further considerations about the solutions of our non-linear, new equation
and the interesting consequences of the present formalism are in preparation
and will appear elsewhere.\\
\h At last, we mentioned at the beginning of this note about
further methods for introducing an helical motion as the classical limit of
the ``spin motion".  We want here to show  how to represent in the
Clifford formalism the {\em extrinsic curvature} approach,$^{16,17}$
due to its possible interest for the development
of the present work.\\
\h Let us first recall that in classical differential geometry one
defines$^{23}$ the Frenet frame $\{e_{\mu}\}$ of a non-null curve $\sig$ by
the so--called Frenet equations, which with respect to proper time $\tau$
write [besides \ $\dox = e_0 = v$]:

\

\hfill{$\ddot{x} = {\dot e}_0 = K_1 e^1 \, ; \;\;
{\dot e}_1 = -K_1 e^0 + K_2 e^2 \, ; \;\; {\dot e}_2 = -K_2 e^1 + K_3 e^3 \,
; \;\; {\dot e}_3 = -K_3 e^2$
\hfill} (22)

\

where the i-th curvatures $K_{\rm i}$ (i=1,2,3) are scalar functions chosen
in such a way that \ ${e_{\rm j}}^2 = -1$, \ with j=1,2,3. \ Quantity $K_1$
is often called curvature, and $K_2, \; K_3$ torsions (recall that in the
3--dimensional space one meets only $K_1$ and $K_2$, called curvature and
torsion, respectively). \ Inserting eqs.(22) into eq.(20), we get for the
Darboux (angular--velocity) bivector:

\

\hfill{$\Om \; = \; K_1 e^1 e^0 \, + \, K_2 e^2 e^1 \, + \, K_3 e^3 e^2 \; ,$
\hfill} (23)

\

so that one can build the following scalar function

\

\hfill{$\Om \cdot \Om \; = \; K_{1}^2 - K_{2}^2 - K_{3}^2 \; = \;
({\dot e}_{\mu} \wedge e^{\mu}) \cdot ({\dot e}_{\nu} \wedge e^{\nu}) \;\; .$
\hfill} (23')

\

\h At this point, one may notice that the square, $K^2$, of the ``extrinsic
curvature" entering Pav\v{s}i\v{c}'s lagrangian is equal to $-K_{1}^2$, \ so
that such a lagrangian$^{16}$ results ---after the present analysis--- to take
advantage only of the first part,

\

\hfill{${\ddot x}^2 \; = \; {\dot e}_0^2 \; = \; -K_{1}^2 \; ,$
\hfill}

\

of the Lorentz invariant (23').
On the contrary, in our formalism the whole invariant $\Om \cdot \Om$
suggests itself as the suitable, complete lagrangian for the problem at issue;
and in future work we shall exploit it, in particular comparing the expected
results with Plyushchay's.$^{37}$\\
\h For the moment, let us stress here only the possibly important result
that the lagrangian \ $\Lc \, = \, \Om \cdot \Om$ does coincide (factors
apart)  along the
particle world--line $\sig$ with the auto--interaction term$^{38}$

\

\hfill{${\theta^5} \; (\drm \theta^{\mu} \wedge \theta_{\mu}) \cdot (\drm
\theta^{\nu} \wedge \theta_{\nu}) $
\hfill}

\

of the {\em Einstein--Hilbert lagrangian density} written (in the Clifford
bundle formalism) in terms of tetrads of 1-form fields $\theta^{\mu}$. \
Quantity $\theta^5 \equiv \th^0 \th^1 \th^2 \th^3$ is the volume element.\\
\h Finally, we can examine within our formalism the third approach: that one
utilizing Grassmann variables.$^{19}$  For instance, if we recall that the
Grassmann product is nothing but the external part \ $A_{\rm r} \wedge
B_{\rm s} \, = \, \lan A_{\rm r} B_{\rm s} \ran_{|{\rm r} - {\rm s}|}$ \
of the Clifford product (where $A_{\rm r}$, $B_{\rm s}$ are a r-vector
and a s-vector, respectively), then the Ikemori lagrangian$^{19}$ can be
immediately translated into the Clifford language and shown to be
equivalent to the BZ lagrangian, apart from the constraint $p^2 = m^2$.

\vfill
\newpage

{\bf 6. ACKNOWLEDGEMENTS}\\

The authors express their gratitude to the Referee, for his
comments and for many references to previous related work; and
acknowledge the kind collaboration of J. Vaz.  They are also
grateful to
Asim O. Barut, A. Campolattaro, R.H.A. Farias, E. Giannetto,
Cesare Lattes, L. Lo Monaco, P. Lounesto, A. Insolia, E. Majorana jr.,
J.E. Maiorino, R.L. Monaco, E.C. de Oliveira, M. Sambataro, S. Sambataro,
Q.G. de Souza  and M.T. Vasconselos for stimulating discussions. \
Two of them
(M.P. and W.A.R.) thank moreover I.N.F.N.--Sezione di Catania, Catania,
Italy, for kind hospitality during the preparation of the first draft of
this work; while F.R. acknowledges a grant from the Fondazione
Bonino--Pulejo, Messina.

\end{document}